\documentclass{PoS}

\usepackage{newtxtext,newtxmath}
\usepackage[T1]{fontenc}
\usepackage{ae,aecompl}
\usepackage{graphicx}	% Including figure files
\usepackage{amsmath}	% Advanced maths commands
\usepackage{amssymb}	% Extra maths symbols
\usepackage{caption}

\title{Simultaneous Fitting of the Spectral Energy Density, Energy-dependent Size, and X-ray Spectral Index vs. Radius of The Young Pulsar Wind Nebula PWN G0.9+0.1}

\ShortTitle{Spectral and Spatial Modelling of PWNe}

\author{\speaker{Carlo van Rensburg} and Christo Venter\\
        Centre for Space Research, North-West University, Potchefstroom Campus, Private Bag X6001, Potchefstroom, South Africa, 2520\\
        E-mail: \email{carlo.rensburg@gmail.com}}

\abstract
{
We have constructed and calibrated a spherically-symmetric, spatially-dependent particle transport and emission code for young pulsar wind nebulae (PWNe). This code predicts the spectral energy distribution (SED) of the radiation spectrum at different positions in a PWN, thus yielding the surface brightness vs. radius and hence the nebular size as function of energy. It also predicts the X-ray spectral index at different radii from the central pulsar, depending on the nebular B-field profile and particle transport properties. We apply the code to PWN G0.9+0.1 and fit these three functions concurrently, thus maximizing the constraining power of the data.  We use a Markov-chain-Monte-Carlo (MCMC) method to find best-fit parameters with accompanying errors. This approach should allow us to better probe the spatial behaviour of the bulk-particle motion, the $B$-field and diffusion coefficient, and break degeneracies between different model parameters. Our model will contribute to interpreting results by the future Cherenkov Telescope Array (CTA) that will yield many more discoveries plus morphological details of very-high-energy Galactic PWNe.
}

\FullConference{6th Annual Conference on High Energy Astrophysics in Southern Africa\\
		1-3 August, 2018\\
		Stonehenge in Africa (Parys), South Africa}

\begin{document}

\section{Introduction}
Pulsar wind nebulae (PWNe) are true multi-wavelength objects, observable from the highest $\gamma$-ray energies down to the radio waveband, sometimes exhibiting complex and highly energy-dependent morphologies. PWNe discoveries during the last decade by ground-based Imaging Atmospheric Cherenkov Telescopes (IACTs) have increased the number of known very-high-energy (VHE, $E >$ 100 GeV) $\gamma$-ray sources to about 200\footnote{http://tevcat2.uchicago.edu/}. Currently nearly 40 of these are confirmed pulsar wind nebulae (PWNe) \cite{Hewitt2015}. The \textit{Fermi} Large Area Telescope (LAT) instrument has detected 5 high-energy $\gamma$-ray PWNe and 11 PWN candidates \cite{3FGL2015} with the X-ray to VHE $\gamma$-ray energy range having 85 PWNe or PWN candidates, 71 of which have associated pulsars \cite{Kargaltsev2012}. 

It is anticipated that the future Cherenkov Telescope Array (CTA), with its order-of-magnitude increase in sensitivity and improvement in angular resolution (point spread function (PSF) in the order of 0.05 deg at VHEs), will discover several more (older and fainter) PWNe and reveal many more morphological details. Due to this and other experiments that will yield improved morphological information we need a refined model that can yield more stringent constraints on several model parameters.

Most current codes model the structure of the PWN as a single sphere (i.e., one-zone models) by assuming spherical symmetry, see e.g.,  \cite{VdeJager2007}, \cite{Zhang_2008}, \cite{Tanaka_Takahara_2011}, \cite{Martin2012}, \cite{Torres2014} and \cite{Lu2017}, although time dependence is an important feature. These codes can predict the spectral energy distributions (SEDs) from PWNe but do not give any information regarding the morphology. On the other end, there are magnetohydrodynamic (MHD) codes that can predict the morphology of PWNe in great detail \cite{Bucciantini2014}, but they in turn cannot predict the SED from the PWNe. There are, however, models that follow a hybrid approach (e.g., \cite{Porth2016}): they model the morphology in great detail using an MHD code, and then use a steady-state spectral model to produce the SED. 

Currently there is a void in the modelling landscape. Thus by adding a spatial dimension to an emission code, we are able to predict the position-dependent SED, the energy-dependent size, and this leads to the index of the spectra for different energy bands and different radii from the pulsar. These can then be used to better constrain model parameters and thus probe the PWN physics more deeply.

In this paper we will give a short discussion on the model we use and show results of our PWN model regarding the radiation spectrum, sizes of the PWN and the X-ray steepening expected as seen in observations.

\section{The Model}

Full details of the code has been published \cite{CvR2018MNRAS_G09}. We will discuss some of the key features here. A modified transport equation of the form 
\begin{equation}
\begin{split}
\frac{\partial N_{\rm{e}}}{\partial t} =& -\mathbf{V} \cdot (\nabla N_{\rm{e}}) +  \kappa \nabla^2 N_{\rm{e}} + \frac{1}{3}(\nabla \cdot \mathbf{V})\left( \left[\frac{\partial N_{\rm{e}}}{\partial \ln E_{\rm{e}}} \right] - 2N_{\rm{e}} \right)   \\
&+ \frac{\partial }{\partial E}(\dot{E}_{\rm{e,tot}}N_{\rm{e}}) +  Q(\mathbf{r},E_{\rm{e}},t)
\end{split}
\label{eq:transportFIN}
\end{equation} 
is solved, with $N_{\rm{e}} \equiv U_{\rm{E}}(\mathbf{r},E_{\rm{e}},t)$ representing the number of particles per unit energy and volume, \textbf{V} the bulk motion of particles, $\kappa$ the diffusion coefficient, $\dot{E}_{\rm{e,tot}}$ the total energy loss rate, including radiation (synchrotron and inverse Compton) and adiabatic energy losses, and $Q$ is the particle injection spectrum with $r$ the radial dimension. This equation is solved in a spherically symmetric system divided into concentric spheres (zones). The particles are injected using a broken power law. The bulk motion of particles are parametrised according to $V(r) = V_0\left(r/r_0\right)^{\alpha_{\rm{V}}}$, with $\alpha_{\rm{V}}$ the velocity profile parameter and thus only having a radial dependence. The $B$-field inside the PWN is modelled using the following parametrised form that allows the magnetic field to have both a spatial and time dependence:
\begin{equation}
B(r,t) = B_{\rm{age}}\left(\frac{r}{r_0}\right)^{\alpha_{\rm{B}}}\left(\frac{t}{t_{\rm{age}}}\right)^{\beta_{\rm{B}}},
\label{B_Field}
\end{equation}
with $B_{\rm{age}}$ the present-day $B$-field at $r = r_0$ and $t = t_{\rm{age}}$, $t$ the time since the PWN's birth, $t_{\rm{age}}$ the PWN age, and $\alpha_{\rm{B}}$ and $\beta_{\rm{B}}$ the $B$-field parameters. The diffusion of particles is modelled using Bohm-type diffusion of the form $\kappa(E_{\rm{e}}) = \kappa_B E_{\rm{e}}/B(r,t)$, and $\kappa_B = c/3e$ with $e$ denoting the elementary charge. We introduce a scaling factor $\kappa_0 \left(\kappa(E_{\rm{e}})\right)$ that is used later to scale the diffusion during fitting.

We solve equation \eqref{eq:transportFIN} using a DuFort-Frankel numerical scheme that gives the particle spectrum in each concentric sphere (zone). The radiation spectra can then be calculated for each zone by using the particle spectrum, which in turn can be used to do a line-of-sight (LOS) calculation to find the predicted surface brightness. The predicted size of the PWN is determined by determining where the surface brightness has dropped off by two thirds. The SED for each zone is used to calculate the index for a certain energy range for different distances from the embedded pulsar, by fitting a power-law curve to the specific energy range.

\section{Results}
In this paper PWN G0.9+0.1 is used as a case study. To find a best fit to the data, we used the model parameters similar to those of Torres et al.\ \cite{Torres2014} and a summary of these are given in Table~\ref{tbl:T}. We freed four parameters that had a significant influence on the SED and size of the PWN, i.e., the $B$-field, the age of the system, the bulk speed of particles, and the diffusion coefficient normalisation ($\kappa_0$). We assume $\alpha_{\rm{B}} + \alpha_{\rm{V}} =-1$ \cite{CvR2018MNRAS_G09}. Thus in our code, only $\alpha_{\rm{B}}$ is free. We have tested other values for $\alpha_{\rm{B}}$, finding that the newly induced discrepancy in the SED outweighs the improved profile of the spectral index. We therefore fix $\alpha_{\rm{B}} = 0$ in this paper. Futhermore, we fix $\beta_{\rm{B}} =-1.0$ to correspond with the expectation for the time evolution of the magnetic energy density (which is linked to the changing spin-down power and expanding PWN radius; \cite{Torres2014}). We used an MCMC ensemble sampler\footnote{http://dfm.io/emcee/current/} to explore the parameter space for these four parameters and the results can be seen in Figure~\ref{fig:corner}, indicating that these parameters are not degenerate and that their best-fit values have reasonable errors. 

{\centering
\begin{minipage}[b]{0.45\textwidth}
   \centering
   \resizebox{1.1\textwidth}{!}{
   \begin{tabular}{crr}
  \hline
  Model Parameter & Symbol & Best fit \\
  \hline
  \textbf{Free parameters}\\
  Present-day $B$-field & $B(t_{\rm{age}})$ & $14.96^{+0.96}_{-0.89}$ $\mu \rm{G}$\\
  Age of the PWN & $t_{\rm{age}}$ & $3~078^{+111}_{-121}~\rm{yr}$ \\
  Particle bulk motion & $V_0$ & $0.089^{+0.013}_{-0.012}~c$ \\
  Diffusion (in units of Bohm) & $\kappa_0$ & $2.15^{+0.40}_{-0.36}$ \\
  \hline
  \textbf{Fixed parameters}\\
  Initial spin-down power ($10^{38}\rm{erg}$ $\rm{s^{-1}}$)& $L_0$ & 1.44  \\ 
  $B$-field parameter & $\alpha_{\rm{B}}$ & 0.0 \\
  $B$-field parameter & $\beta_{\rm{B}}$ & $-$1.0 \\
  $V$ parameter & $\alpha_{\rm{V}}$ & $-1.0$ \\
  \hline
\end{tabular}}
   \captionof{table}{\label{tbl:T}Best-fit parameters for PWN G0.9+0.1 for concurrently fitting the SED as well as the energy-dependent size of the PWN.}
\end{minipage}
\hspace{0.05\textwidth}
\begin{minipage}[b]{0.55\textwidth}
   \centering
   \includegraphics[width=1.0\textwidth]{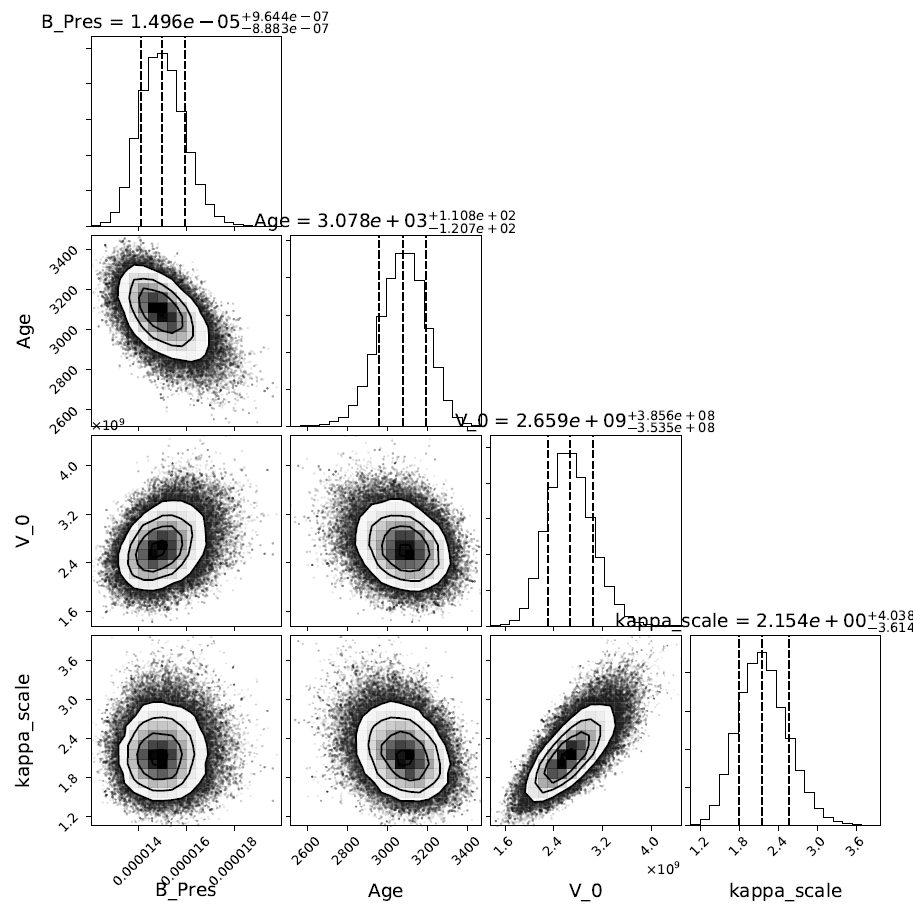}
   \captionof{figure}{\label{fig:corner}MCMC results for the best fit to the data of PWN G0.9+0.1.}
\end{minipage}
}

\begin{figure}[h!]
\includegraphics[width=1.0\textwidth]{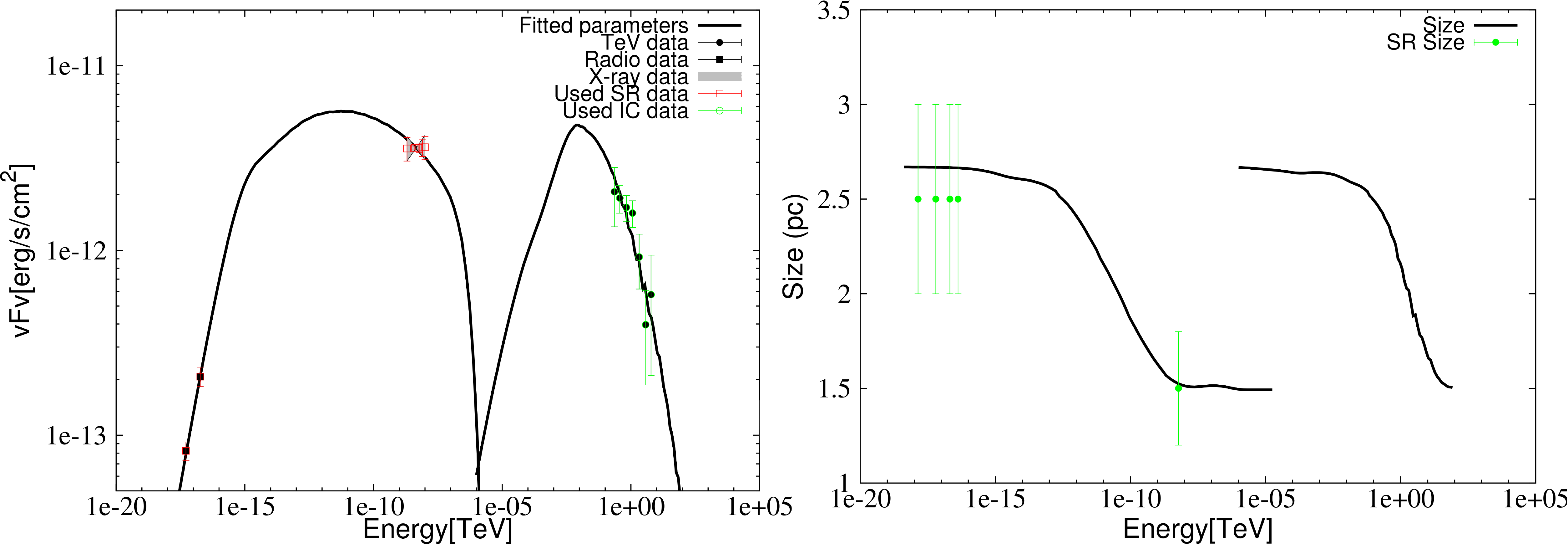}
\caption{\label{fig:SED+sve} \textit{Left:} SED for G0.9+0.1 with added radio \cite{HelfandB1987}, X-ray \cite{Porquet2003} and $\gamma$-ray data \cite{G0.9+0.1_HESS}. \textit{Right:} Size of the PWN as a function of energy for G0.9+0.1 with the observed radio \cite{Dubner2008} and X-ray sizes \cite{Porquet2003}.}
\end{figure}

Figure~\ref{fig:SED+sve} indicates the best fits for the SED in the left panel and the size as a function of energy in the right panel. These results are now possible due to the spatial dependence of our code. These are, however, not the only spatial properties that can be measured. Two other properties frequently measured are the X-ray spectra index and the surface brightness as function of the distance from the central pulsar. 

\begin{figure}[!htb]
    \centering
    \begin{minipage}{.4\textwidth}
        \centering
        \includegraphics[width=1.0\linewidth]{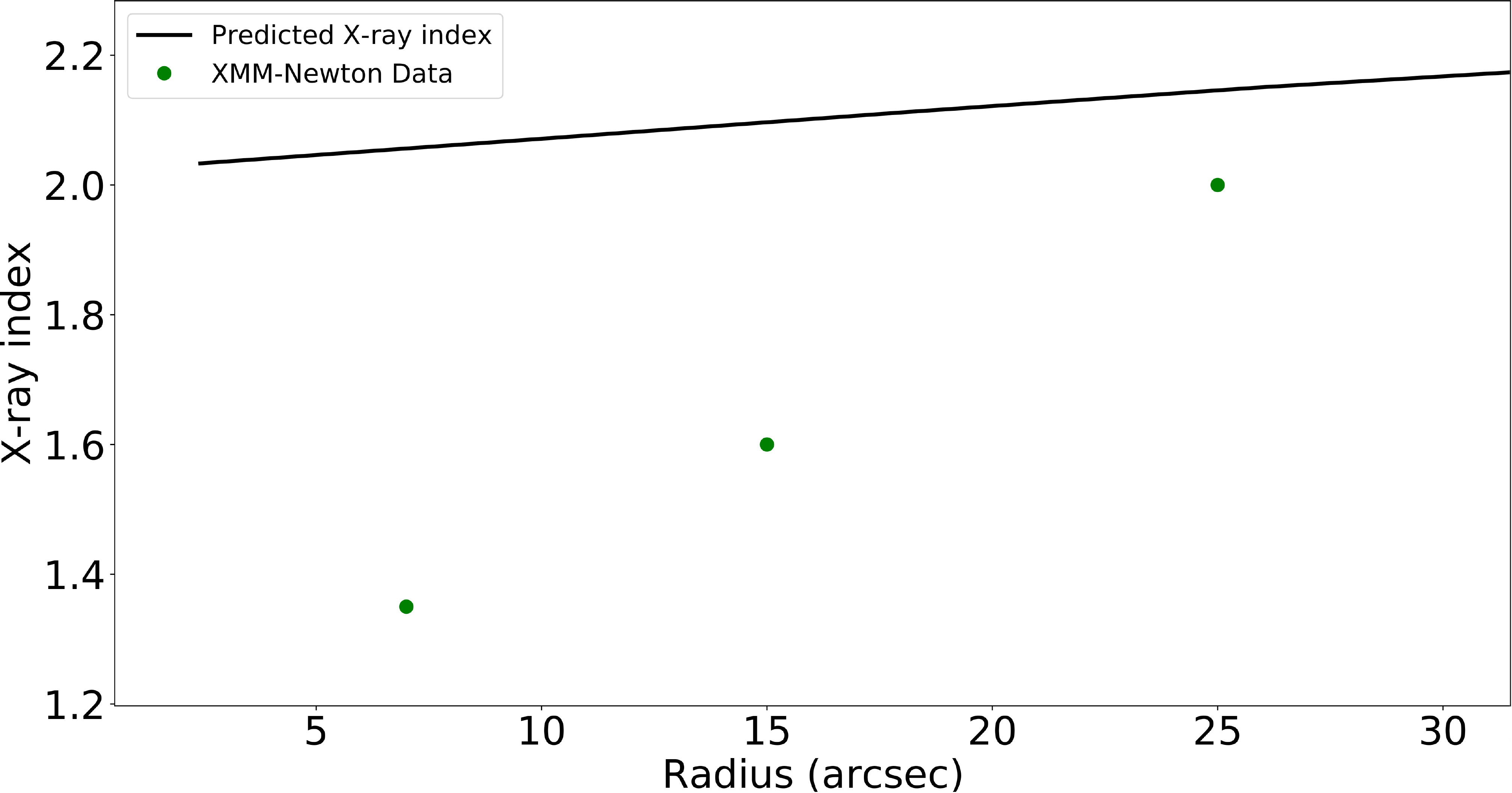}
        \caption{X-ray index as a function of radius for PWN G0.9+0.1 for the parameters as in Table~\ref{tbl:T}.}
        \label{fig:Xray0}
    \end{minipage}%
    \hspace{0.1\textwidth}
    \begin{minipage}{0.4\textwidth}
        \centering
        \includegraphics[width=1.0\linewidth]{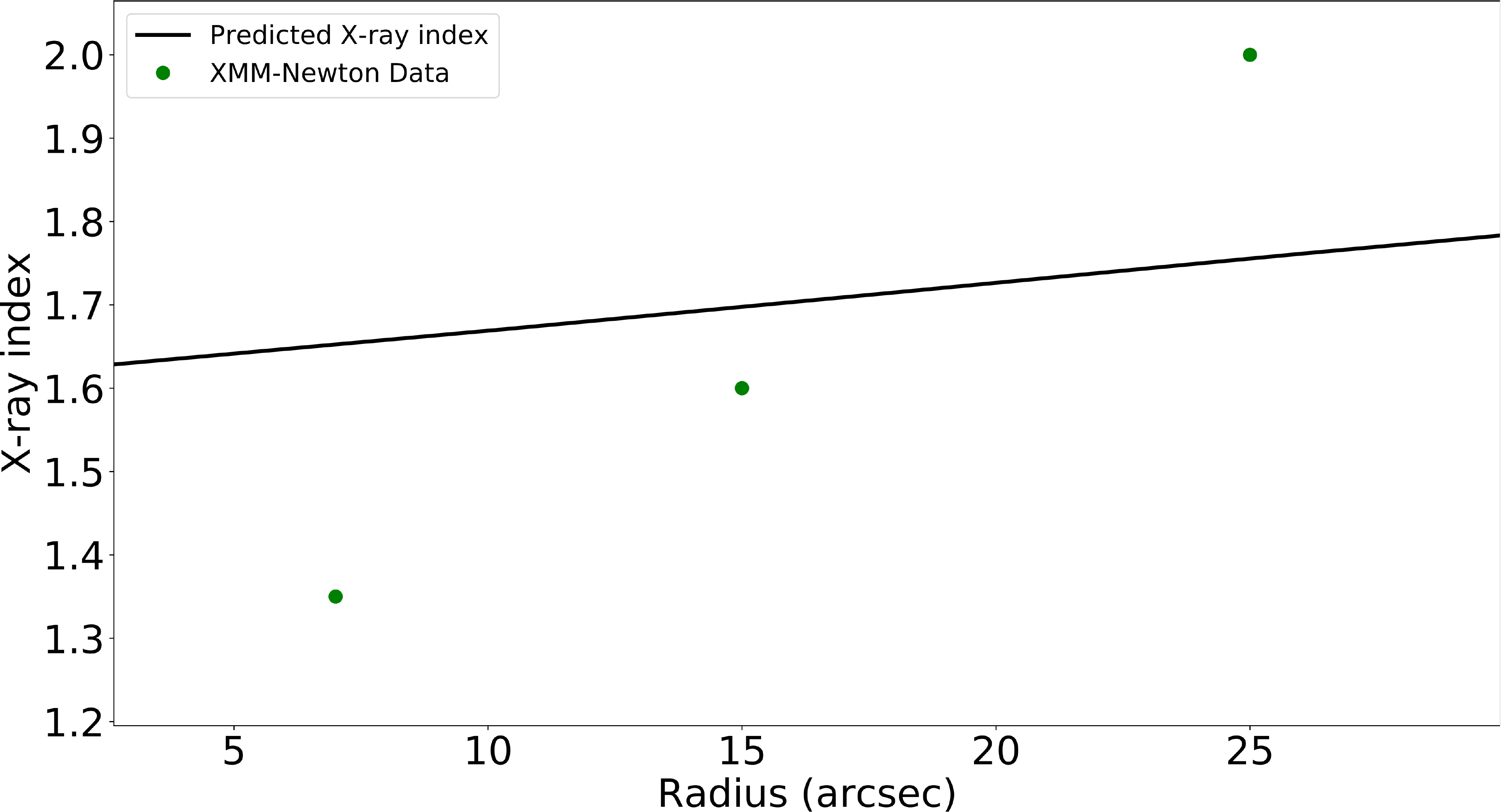}
        \caption{X-ray index as a function of radius for PWN G0.9+0.1 for a change in some of the parameters.}
        \label{fig:Xray1}
    \end{minipage}
\end{figure}

In Figure~\ref{fig:Xray0} the X-ray spectral index is indicated by the black dots measured by the \textit{XMM-Newton} space telescope for the energy range of $2-10$ keV \cite{2012XMMG09}. These clearly exhibit spectral steepening that is an indication of cooling in the PWN. From our code we can use the LOS-integrated SED to fit a power-law function of the following form
\begin{equation}
 E^2\frac{dN}{dE} = kE^{2-\Gamma},
\end{equation}
where $\Gamma$ is the spectral index (black line).

From Figure~\ref{fig:Xray0} we can see that for current model parameters the X-ray index is not predicted correctly. The predicted index do, however, show steepening with $r$. In Figure~\ref{fig:Xray1} the model parameters (e.g., $\alpha_1$ and $\alpha_2$, the indices of the broken power law assumed to represent the particle injection spectrum) were changed by hand, and this caused the predicted X-ray index to represent the data more closely. This, however, changed the predicted SED and the energy-dependent size of the PWN significantly, demonstrating the need for further investigation. These are still preliminary results and a full MCMC best-fit procedure will thus be followed that will include more free parameters to fit all three these observational properties at the same time. 

\section{Conclusions}
We have developed a spherically-symmetric, spatially-dependent particle transport and emission code for young PWNe that is able to fit SEDs and energy-dependent sizes of PWN simultaneously. The code is also able to predict the X-ray index and the surface brightness as function of the distance from the central pulsar. This, however, is still preliminary but in the near future we will use an MCMC best-fit procedure to fit not only the SED and size of the PWN, but also the X-ray index and the surface brightness $-$ all simultaneously.

\section{Acknowledgements}
\footnotesize{
This work is based on the research supported wholly / in part by the National Research Foundation (NRF) of South Africa (Grant Numbers 87613, 90822, 92860, 93278, and 99072). The Grantholder acknowledges that opinions, findings and conclusions or recommendations expressed in any publication generated by the NRF supported research is that of the author(s), and that the NRF accepts no liability whatsoever in this regard.}

%\begin{thebibliography}{99}
\bibliographystyle{JHEP}
\bibliography{Bibliography}

%\end{thebibliography}

\end{document}